\documentstyle[preprint,aps]{revtex}
\begin{document}
\draft
\title{\bf{\LARGE{Time evolution of models described by
one-dimensional discrete nonlinear Schr\"odinger equation}}}
\author{P. K. Datta\cite{mail1} and K. Kundu\cite{mail2} }
\address{Institute of physics, Bhubaneswar - 751 005, India}
\maketitle
\begin{abstract}
The dynamics of models described by a one-dimensional discrete
nonlinear Schr\"odinger equation is studied.  The nonlinearity in
these models appears due to the coupling of the electronic motion to
optical oscillators which are treated in adiabatic approximation.
First, various sizes of nonlinear cluster embedded in an infinite
linear chain are considered. The initial excitation is applied either
at the end-site or at the middle-site of the cluster. In both the
cases we obtain two kinds of transition: (i) a cluster-trapping
transition and (ii) a self-trapping transition.  The dynamics of the
quasiparticle with the end-site initial excitation are found to
exhibit, (i) a sharp self-trapping transition, (ii) an
amplitude-transition in the site-probabilities and (iii) propagating
soliton-like waves in large clusters.  Ballistic propagation is
observed in random nonlinear systems. The
effect of nonlinear impurities on the superdiffusive behavior of
random-dimer model is also studied.
\end{abstract} 

\vspace{.1in}

\narrowtext
\newpage

\section{Introduction}
Strong interaction with the lattice vibrations is one of the basic
mechanisms influencing the transport of quasiparticles such as
electrons or exitons in solids. The consequences have been
investigated employing different methods.\cite{1} The recent approach
to this problem is based on nonlinear equations.
\cite{7,14,8}
One of the simple models with varieties of applications in different
areas is one-dimensional discrete nonlinear Schr\"odinger equation:
\cite{15,16,17,18,19,20,24,25}
\begin{equation}
i \frac{dc_m}{dt} = V (c_{m+1} + c_{m-1}) + (\epsilon_m - \chi_m
|c_m|^2) c_m. \label{1}
\end{equation}
Here $c_m(t)$ is the probability amplitude of the quasiparticle at
site $m$ at time $t$, $V$ is the nearest-neighbor transfer matrix
element, $\epsilon_m$ and $\chi_m$ are on-site energy and
nonlinearity strength of the $m$-th site respectively. Without any
loss of generality we assume $V = 1$. The Eq. (\ref{1}) arises in the
general problem of polaron formation due to the coupling of
quasiparticles with optical oscillators in adiabatic approximation.
The simple form of the Eq.  (\ref{1}) with $\epsilon_m = 0$ and
$\chi_m = \chi$ for all $m$ has been studied numerically since long
back.\cite{15} However, for a two-site system which is called the nonlinear
adiabatic quantum dimer, the self-trapping transition
\cite{8,15} occurs at a critical value of nonlinearity
arbitrary initial conditions.\cite{16,17,18,19,24}  
The applications of the nonlinear
dimer analysis have been made to several experimental situations.
They are, neutron scattering off hydrogen atoms trapped at the
impurity sites in metals,\cite{17} fluorescence
depolarization,\cite{19} muon spin relaxation,\cite{24} nonlinear
optical response of superlattices \cite{29} etc.  The
self-trapping transition also occurs in the extended nonlinear
systems.\cite{25} A possible application is the trapping of hydrogen
ions around the oxygen atoms in metal hydrides.\cite{30} All these
studies have been performed for finite number of nonlinear sites by
assuming that the quasiparticle is localized within the nonlinear
sites.  However, the effect of nonlinear sites embedded in a host
lattice on the dynamics of quasiparticles has been hardly studied in
spite of its importance in real systems. Dunlap et al
\cite{31}
studied the self-trapping transition at a single nonlinear impurity
embedded in a host lattice. 
Chen et al \cite{32} studied the time averaged probability at
the initial occupation site in an infinite linear chain containing
one or many nonlinear impurities. In this study the adiabatic
assumption has been removed. They have also studied the adiabatic
case albeit not in details. So, in this paper we plan to study first
the dynamics of a quasiparticle in an infinite linear chain
containing adiabatic Holstein type impurities.\cite{1} We use two
different kinds of initial conditions. The initial excitation is
applied either at the end-site or at the middle-site of the cluster of the
impurities.

If we consider randomness in site energies $\epsilon_m$ and $\chi_m =
0$ for all $m$ in Eq. (\ref{1}) Anderson theory
\cite{33} predicts that the particle will remain localized within a finite region of
the chain after a sufficient time. So, it is important to investigate
the dynamics of (i) random nonlinear systems (randomness in the
nonlinearity parameter) and (ii) systems where disorder in site
energies and nonlinearity coexist. Regarding the first question
Molina and Tsironis \cite{34} showed the ballistic propagation of the
untrapped electronic fraction in nonlinear random binary alloy. In
this paper we study the transport properties of completely random
nonlinear systems. Feddersen \cite{35a} has studied the effect of
nonlinearity on the Anderson localization. Shepelyansky
\cite{35} has obtained subdiffusive behavior in on-site energy disordered
systems only when the nonlinearity parameter exceeds a critical
value. It is well known that superdiffusive behavior is obtained in
the random-dimer model (RDM).\cite{36} The RDM is characterized by a
set of nonscattered states around the dimer energy. This leads to the
superdiffusive behavior of the mean square displacement of a
particle. We study here the effect of nonlinearity on the
superdiffusive behavior of the RDM.

The organization of the paper is as follows. In Sec. II, we study the
dynamics of the quasiparticle in a cluster of nonlinear sites
embedded in a lattice.  The initial excitation is applied either at the
end-site or at the middle-site of the cluster.  In Sec. III, we
study the dynamics of different kinds of random systems. We end this
article by summarizing our main results.

\section{CLUSTER OF NONLINEAR IMPURITIES EMBEDDED IN A LATTICE}

\subsection{Initial excitation at the end of the cluster}

We consider a system containing a cluster of $n$ number of nonlinear
impurity sites of equal strength \(\chi\) embedded in a host lattice.
All the site energies are assumed to be zero. The initial excitation
is applied at left end-site of the nonlinear cluster.  We call this
zeroth site. The sites on the left and the right of the initial
occupation site are numbered as $m = -1, -2, -3, \cdots$ and $m = 1,
2, 3, \cdots$ respectively. We first study here the time averaged
probability of the nonlinear sites. For $m$-th site it is defined as
\begin{equation}
< P_m > = \lim_{T \rightarrow \infty} \frac{1}{T}\int_0^T|c_m(t)|^2
dt
\end{equation}
with \( |c_m(0)|^2 = \delta_{m,0}\). Here, \(|c_m(t)|^2\) is the
probability of the quasiparticle at $m$-th site at time $t$. We solve
the first order coupled nonlinear differential equations numerically
by using 4-th order Runge-Kutte method. The system is taken as
self-expanding lattice to avoid the boundary effect. For time
averaging we have taken $T =200$ with interval $\Delta T = 0.01$. The
accuracy of the numerical integration is checked through total
probability.  Here, we consider the cases for $n=2, 3, 4, 5$ and
asymptotically large value of $n$ ($n
\rightarrow \infty)$. For $n \rightarrow \infty$ we mean that
the system contains two semi-infinite chains. The perfect chain
without any nonlinearity is connected to the other one which is a
perfect nonlinear chain. The initial excitation is applied at the
junction where the nonlinear impurity exists. In Fig.  1 we have
plotted the time averaged probability at the initial occupation site,
$< P_0 >$ as a function of $\chi$ for different values of $n$. In all
these cases we find that $< P_0 >$ starts increasing significantly
from $\chi = \chi_{cr1}$. For $n = 2$ the value of $\chi_{cr1}$ is
$\sim 2.8$. In general the value of $\chi_{cr1}$ increases with
increasing the size of the nonlinear cluster. A sharp transition in
$< P_0 >$ occurs at $\chi_{cr2} \sim 4.23$ for all values of $n$.  In
the region between $\chi_{cr1}$ and $\chi_{cr2}$ we obtain
fluctuations in $< P_0 >$. For a better understanding of this
behavior we study next the time averaged probability of the
unoccupied nonlinear sites.

In Fig. 2 we plotted the time averaged probability of the initially
unoccupied nonlinear site \((< P_1 >)\) of the dimer as a function of
$\chi$. For comparison we have also plotted \(< P_0 >\). When the
nonlinearity strength $\chi$ exceeds $\chi_{cr1}$ we observe both $<
P_0 >$ and $< P_1 >$ increase with increasing $\chi$ and their values
are almost equal. This implies that the particle oscillates with a
finite probability among the dimer sites. This partial localization
or trapping within the cluster can be understood from the following
way. At $t = 0$ the energy level of the sites $m = -1, 0$ and $1$ are
$0, -\chi$ and $0$ respectively.  With increasing the time the
site-probability of $m = 0$ decreases and of $m = 1$ increases.
Consequently, the energy level of the sites $m = 0$ and $1$ moves
upward and downward from the original position respectively. Thus,
the energy gap between the sites $m = 0$ and $1$ becomes smaller than
the gap between $m = -1$ and $0$. So, the initially localized
particle at site $m =0$ favors the nearest-neighbor nonlinear site
(i.e. $m = 1$) and the energy level of that site decreases. At the
same time the energy gap between the sites $m = 1$ and $2$ increases.
Thus the particle feels a quantum well and it oscillates within the
well with a finite probability. Of course, some probability will
escape along the leads in both directions of the dimer. Now the
leakage of the probability through the leads reduces with the
increase of $\chi$ due to the increase in the energy gaps between the
sites $m = 0$ and $-1$ and between $m = 1$ and $2$.  When $\chi$
attains a critical value, say $\chi_{cr1}$, these gaps become
sufficiently large to trap the particle within the cluster.  So,
$\chi_{cr1}$ marks the onset of the cluster-trapping transition of
the particle. With a further increase in $\chi$, the competition
among the nonlinear sites to trap the particle starts.  Depending on
the strength of the nonlinearity the particle is preferentially
trapped either at the initially occupied site or at the unoccupied
site (see the inset of Fig. 2).  When $\chi$ is just below
$\chi_{cr2}$, the value of $< P_1 >$ is much larger than $< P_0 >$.
But, when $\chi$ crosses $\chi_{cr2}$ we find sharp fall in $< P_1 >$
and an increase in $< P_0 >$.  For further increase of $\chi$, $< P_0
>$ increases and $< P_1 >$ decreases gradually. We do not obtain any
further transition. So, $\chi_{cr2}$ is called the critical value of
$chi$ for self-trapping transition.

To understand the behavior in the fluctuation regime of time averaged
site-probabilities we study their temporal behavior. This is shown in
Fig. 3. We find both the site-probabilities oscillate against each
other initially for some time. For a given value of $\chi$, the time
period of the oscillation increases and the amplitude decreases with
increasing time. But after a few oscillations, we observe a
transition where the amplitude of the two oscillations decreases
suddenly and the phase is just opposite to each other (see Fig.
3(a)). The amplitude-transition in the site-probabilities always
occurs simultaneously. Furthermore, the transition occurs from the
peak of the oscillation in one case and from dip in the other one.
Consequently, one of the site-probabilities oscillates with a more
mean probability than the other.  Thenceforth, the amplitude of the
oscillation of the probabilities decreases with time.  This kind of
transition is obtained in the nonadiabatic nonlinear quantum dimer
problem in the presence of rapid vibrational relaxation caused by the
damping in the lattice vibration.\cite{38} Two transitions are
observed, static transition at $\chi = 2 V$ \cite{15} and a dynamic
transition at $\chi = 4 V$.\cite{16} The static transition is
governed by the relaxation term of the lattice vibration. Here, it
seems that the leads at both the ends of the nonlinear dimer
introduce effectively a damping term in the lattice vibration. This
effect appears through the escape probability from the dimer cluster.
However, the main difference here is that the transition occurs at
different values of $\chi$. The transition
time as well as the number of the  dynamical
adiabatic dimer type oscillations decreases with increasing $\chi$
(compare Fig.  3(a) and 3(b)).  Furthermore, near $\chi_{cr2}$ the
number of the dynamical adiabatic dimer kind oscillations does not
reduce for a wide range of $\chi$.  Consequently, we do not
find any fluctuation in time averaged
probability in this region (see the region $3.87 < \chi <4.23$
 of the inset of Fig.  
2). The amplitude-transition in the
site-probabilities occurs after half of the period of the oscillation
(see Fig. 3(c)). Thus, just below $\chi_{cr2}$, $< P_1 >$ is found to
be much larger than $< P_0 >$.  When $\chi$ just crosses $\chi_{cr2}$
the transition occurs at time which is even less than the half period
(see Fig. 3(d)).  As there is no dynamical adiabatic dimer type
oscillation we do not find any further transition in the amplitude of
the site-probabilities with the increase of $\chi$.  So, in contrast
to the Ref. \cite{32}, we find that the quasiparticle recognizes both
the nonlinear impurities just above $\chi_{cr1}$ and for further
increase of $\chi$, amplitude-transition of the site-probabilities
occurs.  Furthermore, we obtain damped oscillation in the
site-probabilities for $\chi > \chi_{cr2}$.  This has to be
contrasted with the regular oscillation above $\chi = 4$ in an
isolated nonlinear dimer.\cite{16} We also study the time averaged
probability at the nonlinear sites for $n = 3, 4$ and $5$. In all
these cases we observe the same behavior as observed in the dimer
embedded in a host lattice.

For a relatively large size nonlinear cluster (e.g. $n = 30$) the
partial localization of the particle is found to occur at different
region of the cluster. We study the particle propagation in a lattice
for the case of $n = 30$ for different values of $\chi$.  For small
value of $\chi$ ($\chi \le 3.1$) 
we obtain the delocalization behavior of the particle.  For further
increase of $\chi$ we find a soliton-like wave,
extended over a few sites, moves through the nonlinear
cluster.  Thus, we obtain an oscillation of the soliton-like wave in
the cluster. This is shown for $\chi = 3.51$ in Fig. 4. It should be
noted that the maximum probability is 
found within this soliton-like wave.  The time period of the
oscillation of the wave within the cluster is increased with time.
This indicates that if we increase the time, the oscillation will
cease and the soliton-like wave will be localized within a few sites
of the nonlinear cluster.  This is exactly obtained in Fig. 4. 
The number of oscillation decreases before
the localization with increasing $\chi$. The position of the
localization of the wave, however, depends on the value of $\chi$ but
the pattern is not discerned here. For further increase of $\chi$ we
do not find any oscillation. It moves along the cluster and gets
trapped within a few sites of nonlinear cluster.  With increasing the
value of $\chi$ the trapping region in general moves towards the
initial excitation site and we obtain a sharp self-trapping
transition at $\chi_{cr2} \sim 4.23$.  It should be noted that the
width and peak-value of the soliton-like wave decreases and increases
respectively with increasing the value of $\chi$. Though we are not
able to probe the value of $\chi$ where the soliton-like wave starts
to form but it is within $\chi = 3.1$ and $3.2$.  It should also be
noted that the movement of the soliton-like wave is obtained only
when the size of the nonlinear cluster is much larger than the width
of the wave.

In asymptotic limit (i.e. $n \rightarrow \infty$) we also obtain 
the formation of soliton-like wave extended over a few sites in the
nonlinear cluster in the system.  Here, we do not find any
oscillation of the wave as obtained in the case of $n = 30$. For
lower values of $\chi$ the wave moves along the lattice but finally
it is trapped in a region as shown in Fig. 5. If we increase the
value of $\chi$ the localization regime of the wave moves towards the
initially occupied site and thus obtain a sharp self-trapping
transition at $\chi_{cr2} = 4.23$. The width and the peak-value of the wave
decreases and increases respectively with increasing $\chi$. 
However, we did not
probe the region of $\chi$ where the soliton-like wave starts to
form.

\subsection{Initial excitation at the middle of the cluster}

We study here the same system but the initial excitation is given at
the middle site $(m = 0)$ of the cluster. The cluster contains odd
number of sites. As the system is symmetric around the initial
occupation site we do not find the asymmetric probability
distribution.  So, the properties in this system should be different
from the earlier cases. The time averaged probability at the initial
excitation site is shown in Fig. 6 for different values $\chi$ and
$n$. In case of single nonlinear impurity system we obtain the
self-trapping transition at $\chi_{cr1} \sim 3.2$.\cite{31,32} For
higher values of $n$ we find $< P_0 >$ increases significantly from $
\chi = \chi_{cr1}$ and it
characterizes the cluster-trapping transition. The value of
$\chi_{cr1}$ for $n = 3$ is $\sim 2.4$ which is much less than the
self-trapping transition value of $\chi$ for $n = 1$. The value of
$\chi_{cr1}$ in general increases with increasing the size of
cluster. In asymptotic limit (i.e. $n \rightarrow \infty$) the value
of the transition
point is  $\chi_{cr1}^{asy} \sim 3.5$.\cite{39} We further study the
time averaged probability of the neighboring nonlinear sites of the
zeroth site for $n = 5$ (see Fig.  7).  We find that the time
averaged probability of the other nonlinear sites also starts
increasing from $\chi_{cr1}$. The value of $< P_m >$ decreases as we
go away from the initial occupation site i.e. as $| m |$ increases.
This means that beyond $\chi_{cr1}$ the particle lies within a few
sites of the cluster with center at the initial excitation site.
Again, as both sides of the zeroth site contains nonlinearity the
particle is attracted by the nonlinear sites in both directions.
Thus, with increasing the size of the cluster we find that $< P_0 >$
decreases and consequently the value of $\chi_{cr1}$ increases.
Beyond $\chi_{cr1}$ we find that the time averaged probability of the
nonlinear sites (except the zeroth site) first increases and then
decreases with increasing the value of $\chi$. But $< P_0 >$
gradually increases with increasing the value of $\chi$. Thus, we
obtain the localization of the particle at $m = 0$ with maximum
probability which is called self-trapping. However, in this case we
do not obtain any sharp self-trapping transition as seen in the
previous case. In the study of particle propagation we observe
localized soliton-like wave with the peak-value at $m = 0$.  The
width and peak-value of the wave decreases and increases respectively
with increasing the value of $\chi$.  This is obtained in all cases
discussed here.

\section{RANDOM SYSTEMS}

The ballistic motion of an initially localized particle in a
one-dimensional nonlinear random binary alloy has been observed
recently.\cite{34} Here we also show the ballistic motion of a
particle in completely random nonlinear systems. The random nonlinear
systems are characterized by random distribution of the nonlinearity
parameter, $\chi$ with the values $0 < \chi_m < \chi_{max}$. All the
site-energies are assumed to be zero. The initial
excitation is applied at the zeroth site. Furthermore, we assume that
the value of $\chi_0$ is $\chi_{max}$. The MSD is defined as
\begin{equation}
< m^2 > = \sum_{m = -\infty}^{\infty} m^2 |c_m(t)|^2
\end{equation}
with the initial condition $|c_m(0)|^2 = \delta_{m,0}$. After some
initial transient behavior the speed $(\sqrt{<m^2>}/t)$ of the
particle is setteled down to a constant which depends on $\chi_{max}$.
The speed of the particle decreases with increasing $\chi_{max}$. Around
the critical value of $\chi_{max}$ the speed decreases drastically and
beyond this region we observe the slow decay of the speed. The
untrapped portion of the probability leads to the ballistic motion of
the particle above the critical value of $\chi_{max}$.\cite{34}
To obtain the critical value of $\chi_{max}$ we have plotted the time
 averaged
probability of the initial excitation site for different realizations
in Fig. 8. The critical value of $\chi_{max}$ are found to be in the
range of $\sim 3$ and $\sim 3.5$.
Beyond the critical value, $< P_0 >$ in general increases very fast
with increasing $\chi$ but with a certain degree of sample to sample
variation. Within the region of $\chi
\sim 3$ and $\chi \sim 4.5$ we obtain a large
deviation in $< P_0 >$ for different realizations. Beyond this region
$< P_0 >$ increases slowly as $\chi_{max}$ goes up and the
probability of all other sites decreases. In this limit the random
nonlinear lattice to a good approximation can be replaced by a
perfect lattice with a single nonlinear defect of same strength
($\chi_{max}$). This is also true for perfect nonlinear system. 
It can be shown that beyond the transition region of $\chi$ the speed of the
particle in random nonlinear system, perfect nonlinear system and the single
nonlinear impurity problem are almost equal. We study
next the effect of nonlinearity in the superdiffusive motion of the
random-dimer model (RDM).

The RDM is a binary alloy containing two types of atoms with site
energies $\epsilon_a$ and $\epsilon_b$. The restriction on the
randomness is that the site energy $\epsilon_a$ appears in a pair
which is called a dimer. So, the system is the random distribution of
the dimer and other site energy $\epsilon_b$. If we assume
$\epsilon_b = 0$ and $V = 1$ then we obtain $\sim \sqrt{N}$ number of
nonscattered states around $\epsilon_a$.\cite{40} Here, $N$ is the
length of the sample. The MSD goes as $< m^2 > \sim t^{3/2}$.
\cite{36} This is obtained only when
$|\epsilon_a| < 2 $. We now study the effect of nonlinearity on the
transport properties of the RDM.  We assume that all the sites have
equal nonlinearity strength and it is $\chi$.  It should be noted
that by adding the nonlinearity the effective site energies are
altered. Thus initially the dimer correlation is distorted. As the
system contains escape probability, the distortion of the dimer
correlation decreases with increasing time. So, we expect the
superdiffusive behavior in the MSD. This is exactly obtained (see Fig.
9). The exponent of MSD is $\sim 1.5$. We also study the time
averaged probability of the initial excitation site which is one of
the dimer sites.  We obtain a sharp self-trapping transition at
different values of $\chi_{cr2}$ which increases as $\epsilon_a$
increases.  Thus, by increasing the site energy we can alter the value
of $\chi_{cr2}$.  For negative values of $\epsilon_a$ the value of
$\chi_{cr2}$ remains almost constant. If the initial excitation is
applied at the site where dimer is absent we obtain the opposite
behavior. That is, for positive values of $\epsilon_a$ the
$\chi_{cr2}$ almost remain constant and for negative values of
$\epsilon_a$, $\chi_{cr2}$ changes significantly. Above $\chi_{cr2}$ we
also obtain the superdiffusive behavior in MSD.

\section{Summary}

We have studied the dynamics of quasiparticles in different kinds of
nonlinear systems. We first studied the dynamics of the quasiparticle
in a cluster of nonlinear impurities embedded in a perfect linear
host lattice.  The initial excitation is applied either at the
end-site or at the middle-site of the nonlinear cluster. In both the
cases we studied the time averaged site-probabilities and the
particle propagation. In the former case we observed the
cluster-trapping transition at $\chi_{cr1}$ due to the localization
of the quasiparticle within the nonlinear cluster.  For $\chi >
\chi_{cr1}$ the amplitude-transition in the probabilities of the
nonlinear sites is obtained showing that escape probability through
linear sites is analogous to a damping term in oscillator equation of
motion.  The absence of any well defined transition in the amplitude
of the site-probabilities beyond $\chi_{cr2}$ indicates the
self-trapping at this value of $\chi$.  This value of $\chi_{cr2}$ is
$\sim 4.23$ for any size of the cluster of the nonlinear sites. This
clearly indicates that a single mechanism is responsible for
self-trapping.  For relatively large size cluster we observed that
the localization of the quasiparticle occurs in the cluster in the
form of soliton-like wave extending over a few sites.  This
cluster-localization starts between $\chi =3.1$ and $3.2$.  In the
asymptotically large size of the nonlinear cluster we also find that
soliton-like wave moves along the cluster but after some time it
localizes in a region. This localization time and the trapping region
depends on $\chi$.  When the initial excitation is applied at the
middle site of the cluster, the cluster-localization also occurs. The
critical value in general increases with increasing the size of the
cluster. However, here we do not find any sharp self-trapping
transition as well as the transition in the amplitude of
site-probabilities.

The MSD in random nonlinear system shows the ballistic motion
and the speed decreases significantly in the transition regime. 
The initial excitation is applied in this case at the site
with maximum nonlinearity. The time averaged probability of the
initial excitation site is also studied for different realizations.
In each cases we obtained rapid increase in $< P_0 >$ within a range
of $\chi$.  It should be noted that beyond this region of $\chi$ the
dynamics of the quasiparticle in the single nonlinear impurity
system, in random nonlinear systems and in the perfect nonlinear
system are similar.  The effect of nonlinearity on the superdiffusive
behavior of the RDM has also been studied. The exponent of the MSD
remains almost same. With increasing the strength of the nonlinearity
the prefactor of the MSD decreases with the increase in $\chi$. The
self-trapping transition also occurs in this case. The self-trapping
value of $\chi$ increases with increasing $\epsilon_a$ provided
$\chi$ and $\epsilon_a$ both are positive quantity. For negative
values of $\epsilon_a$, this value of $\chi$ remains almost constant
with increasing $\chi$. This aspect will be studied latter.

\begin{figure}
\caption{Time averaged probability at the initial excitation site $(<
P_0>)$ as a function of $\chi$ for different size $(n)$ of nonlinear
cluster. The initial excitation is applied at the end-site $(m = 0)$
of the cluster.}
\end{figure}

\begin{figure}
\caption{Time averaged probability of initially occupied $(< P_0 >)$ and
unoccupied site $(< P_1 >)$ of the nonlinear dimer embedded in a host
lattice as a function of $\chi$. The inset shows details of the $<
P_0 >$ and $< P_1 >$ with $\chi$ in the fluctuation regime. The
initial excitation is applied at the end-site $(m = 0)$ of the
dimer.}
\end{figure}

\begin{figure}
\caption{Plot of site-probabilities of the dimer sites embedded in a
host lattice as a function of time $(t)$ for different values of
nonlinearity parameter $\chi$ as follows: (a) $\chi = 3.72$; (b)
$\chi = 3.77$; (c) $\chi = 4$ and (d) $\chi = 4.4$. In all these
cases the initial excitation is applied at the end-site $(m = 0)$ of
the dimer.}
\end{figure}

\begin{figure}
\caption{Electronic probability propagation profile as a function of
time $(t)$ for $\chi = 3.51$. Here, $n = 30$ and the initial
excitation is applied at the end-site
$(m = 0)$ of the nonlinear cluster.}
\end{figure}

\begin{figure}
\caption{Same as Fig. 4 but $n \rightarrow \infty$ and $\chi = 3.6$.}
\end{figure}

\begin{figure}
\caption{Time averaged probability of the initial excitation site
$(< P_0 >)$ as a function of $\chi$ for different values of $n$. The
initial excitation is applied at the middle-site $(m = 0)$ of the
nonlinear cluster.}
\end{figure}

\begin{figure}
\caption{Time averaged probability $(< P_m >)$ of the nonlinear sites
of the cluster of size $n = 5$ embedded in a host lattice as a
function of $\chi$. The initial excitation is applied at the
middle-site $(m = 0)$ of the cluster. As the system is symmetric
around the zeroth site $< P_{-1} > = < P_1 >$ and $< P_{-2} > = < P_2
>$.}
\end{figure}

\begin{figure}
\caption{Time averaged probability of the initial excitation site
$(< P_0 >)$ as a function of $\chi_{max}$ for different realizations
of random nonlinear system.}
\end{figure}

\begin{figure}
\caption{Plot of $< m^2 >/t^{3/2}$ as a function of time $(t)$ for
the RDM with different values of $\chi$. Here, $\epsilon_a = 1$.}
\end{figure}

\end{document}